\pdfoutput=1
\documentclass[sigconf]{acmart}

\usepackage{cite}
\usepackage{amsmath,amssymb,amsfonts}
\usepackage{multirow}
\usepackage{multicol}
\usepackage{algorithmic}
\usepackage{bigstrut}
\usepackage{graphicx}
\usepackage{textcomp}
\usepackage{array}
\usepackage{booktabs}
\usepackage{xcolor}
\usepackage{algorithm}
\usepackage{algorithmic}

\newcommand{\squishlist}{
\begin{list}{$\bullet$}
 { \setlength{\itemsep}{0pt}      \setlength{\parsep}{0pt}
   \setlength{\topsep}{0pt}       \setlength{\partopsep}{0pt}
   \setlength{\listparindent}{-2pt}
   \setlength{\itemindent}{-5pt}
   \setlength{\leftmargin}{1em} \setlength{\labelwidth}{0em}
   \setlength{\labelsep}{0.5em} } }
\newcommand{\squishend}{
    \end{list}  }

\renewcommand\footnotetextcopyrightpermission[1]{}
\def\BibTeX{{\rm B\kern-.05em{\sc i\kern-.025em b}\kern-.08emT\kern-.1667em\lower.7ex\hbox{E}\kern-.125emX}}
    
\begin{document}

%
\title{Enabling Efficient and Flexible FPGA Virtualization\\for Deep Learning in the Cloud}

\author{Shulin Zeng, Guohao Dai, Hanbo Sun, Kai Zhong\\ Guangjun Ge, Kaiyuan Guo, Yu Wang, Huazhong Yang}
\email{zengsl18@mails.tsinghua.edu.cn, yu-wang@tsinghua.edu.cn}
\affiliation{%
  \institution{Tsinghua University, Beijing, China}
}

%

%

%
\begin{abstract}



FPGAs have shown great potential in providing low-latency and energy-efficient solutions for deep neural network (DNN) inference applications. Currently, the majority of FPGA-based DNN accelerators in the cloud run in a time-division multiplexing way for multiple users sharing a single FPGA, and require re-compilation with $\sim$100 s overhead. Such designs lead to poor isolation and heavy performance loss for multiple users, which are far away from providing efficient and flexible FPGA virtualization for neither public nor private cloud scenarios.

To solve these problems, we introduce a novel virtualization framework for instruction architecture set (ISA) based on DNN accelerators by sharing a single FPGA. We enable the isolation by introducing a two-level instruction dispatch module and a multi-core based hardware resources pool. Such designs provide isolated and runtime-programmable hardware resources, further leading to performance isolation for multiple users. On the other hand, to overcome the heavy re-compilation overheads, we propose a tiling-based instruction frame package design and two-stage static-dynamic compilation. Only the light-weight runtime information is re-compiled with $\sim$1 ms overhead, thus the performance is guaranteed for the private cloud. Our extensive experimental results\footnote{We provide several available demonstrations for our FPGA virtualization solution on Aliyun f3: https://github.com/annoysss123/FPGA-Virt-Exp-on-Aliyun-f3} show that the proposed virtualization design achieves 1.07-1.69x and 1.88-3.12x throughput improvement over previous static designs using the single-core and the multi-core architectures, respectively.


\end{abstract}

%

%
\maketitle

\section{Introduction}

As we are now in the artificial intelligence (AI) era, deep learning~\citep{lecun2015deep} is now playing a more important role in various domains~\citep{alexnet, resnet, amodei2016deep, conneau2016very}. Deep neural network (DNN) inference tasks take up the majority of the total deep learning workloads in the cloud. According to the report of Facebook, data analysis demands based on inference tasks is doubling each year in the data center~\citep{park2018deep}. Moreover, in the data center of Amazon, inference tasks make up nearly 90\% of the total deep learning tasks~\citep{aws}. 


Due to the advantages of programmability, high performance, and energy efficiency, many cloud vendors have provided their cloud services using FPGAs in recent years, such as Amazon~\citep{amazon2019aws}, Alibaba~\citep{alibaba2019f1}, and Microsoft~\citep{firestone2018azure}. FPGA accelerators provide high energy efficiency and performance solutions for DNN inference tasks compared with GPUs, which has been well-researched in previous work~\citep{han2017ese, guo2017angel}. However, most FPGA based deep learning accelerators are optimized to provide the optimal performance for single-task and static-workload scenarios~\citep{chen2019cloud, aydonat2017opencl, xilinx2018xDNN, fowers2018configurable}, which are hard to meet the requirement of cloud computing.


Virtualized FPGA based DNN accelerators are needed to support multi-client and dynamic-workload scenarios in the data center. As Nvidia provides virtualized GPUs for flexible deployment in the data center to support every workload, with the features of live migration, optimal management and monitoring, and multi-GPU in a single virtual machine~\citep{nvidia-virtual}, virtualization is the most basic feature of cloud computing in the data center. Besides, Michael \emph{et al.}~\citep{michael2019performance} demonstrate that the virtualized GPU can obtain better DNN inference performance when using half the resources than when using all the resources, meaning that virtualization can improve the system utilization by dynamically sharing and allocating hardware resources. Moreover, Uta \emph{et al.}~\citep{uta2018performance} show that most big data applications have dynamic-workload feature and resources variability, leading to increased demand for virtualization. 

\begin{figure*}[!tp]
\centering
\includegraphics[width=17.5cm]{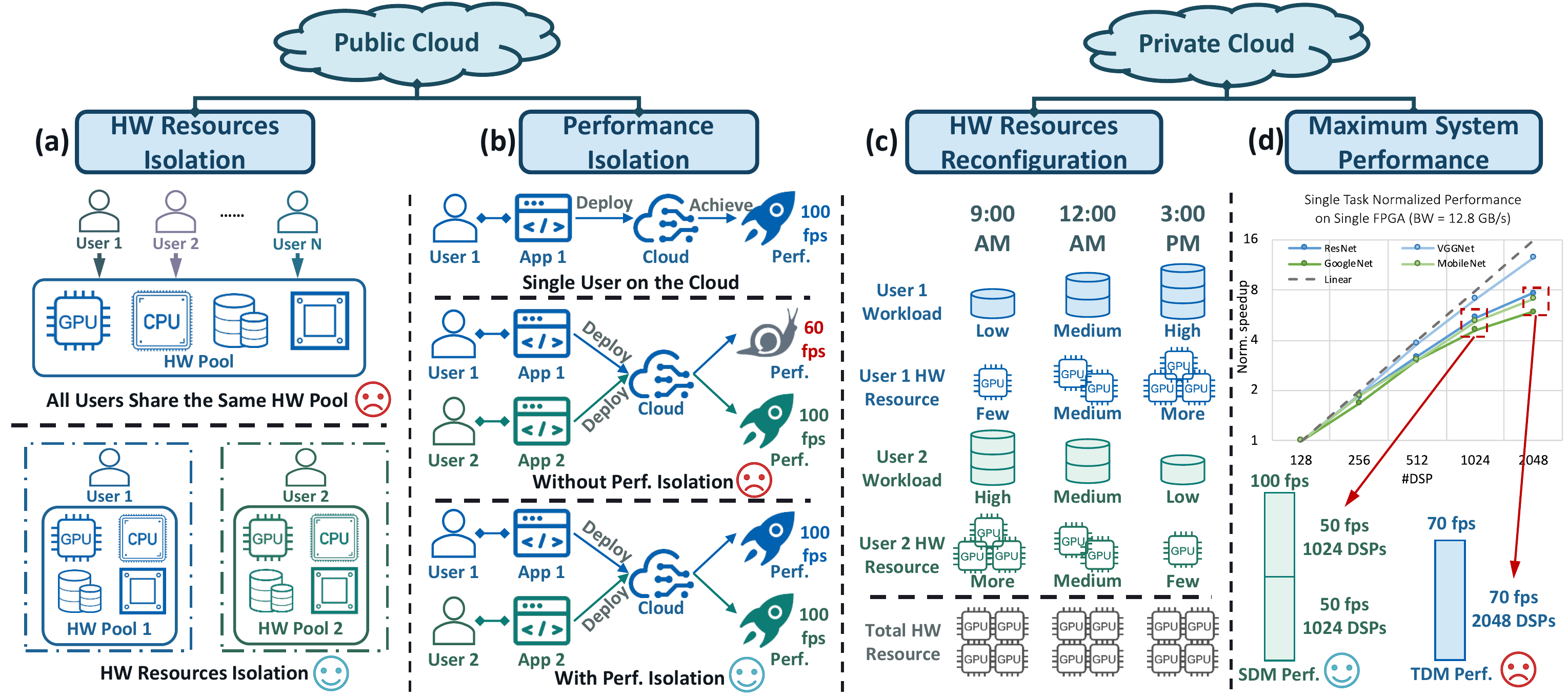}
\vspace{-10pt}
\caption{From left to right: (a) Hardware resources isolation in public cloud. (b) Performance isolation in public cloud. (c) Hardware resources reconfiguration in private cloud. (d) Maximum system performance in private cloud.}
\vspace{-10pt}
\label{fig:motivation-new}
\end{figure*}

As shown in Figure~\ref{fig:motivation-new}, there are two typical scenarios in the cloud: the public cloud and the private cloud. The key idea of virtualization in the first scenario is to ensure physical resources and performance isolation between the users. Physical resources isolation is an important requirement to guarantee the security of both the users and cloud vendors. Performance isolation means that the performance of each user should not be disturbed by the concurrent execution of multiple users. As for the second scenario, virtualization needs to provide computing power reconfiguration ability to ensure that the overall system performance can be maximized under different situations with multiple users and dynamic workloads.

There are multiple FPGAs in the data center, with each FPGA as a single node. The scope of this work is to enable virtualization on a single FPGA of the node level. The exploration of our proposed methods on the system multi-node level will be exciting, which we leave as future work.
Currently, there are two methods of sharing a single FPGA: time-division multiplexing (TDM) and space-division multiplexing (SDM). The former hardly need to re-program the FPGA, but schedules multiple tasks on the same physical resources in the form of time slices. Its disadvantage is difficult to achieve physical resources isolation, resulting in poor security. SDM can be easier to achieve better physical and performance isolation, and we demonstrate that our proposed SDM-based multi-core hardware design can achieve the performance isolation of less than 1\% deviation for multiple users in the public cloud scenario.  

However, SDM has the problem of excessive reconfiguration overheads. The computing resources allocated to a user may change at any time due to the dynamic workload in the private cloud, as shown in Figure~\ref{fig:motivation-new}(c). Therefore, each time a hardware resources reconfiguration occurs, template-based DNN accelerators~\citep{zhang2018dnnbuilder} require the FPGA to be reprogrammed, while instruction set architecture (ISA) based DNN accelerators~\citep{guo2017angel} require the instruction files to be regenerated and reloaded based on the re-allocated hardware resources for each user. Both designs require several minutes of reconfiguration overhead~\citep{papadimitriou2011performance, xing2019dnnvm}, which is unacceptable for DNN inference applications that usually run with the latency within the order of milliseconds. Besides, such a heavy reconfiguration overhead will lead to problematic tail latency of serving DNN requests. However, we find an optimization possibility in reconfiguration overhead of ISA-based DNN accelerators by only re-compiling light-weight runtime information. The basic idea is to generate fine-grained instruction packages in advance during the offline deployment, and then integrate and re-allocate the instruction packages according to the allocated hardware resources for each online reconfiguration. In this paper, we propose a two-stage static-dynamic compilation process to reduce the online reconfiguration overhead to $\sim$1ms, while the average single-task performance loss of multi-core sharing is negligible compared to the single-core baseline design.

Moreover, we find that there is a non-linear relationship between the performance and the hardware resources of the ISA-based DNN accelerator in the single-task scenario due to the limited off-chip bandwidth (BW). As shown in Figure~\ref{fig:motivation-new}(d), the single-task throughput is 50 fps and 70 fps with 1024 DSPs, 6.4 GB/s BW and 2048 DSPs, 12.8 GB/s BW, respectively. TDM will run with all DSPs and BW, resulting in the overall throughput of 70fps, while SDM can run with two cores of half DSPs and BW, achieving a better throughput of $50\times2=100$ fps in the multi-task scenario. We will show that our SDM-based multi-core virtualized design can improve the overall system throughput compared to the TDM-based single-core design.

In this paper, we focus on enabling efficient and flexible FPGA virtualization for deep learning inference applications in the cloud by sharing a single FPGA in an SDM manner. To support FPGA virtualization for both public and private cloud scenarios, we introduce a novel FPGA virtualization framework for ISA-based DNN accelerators, including the hardware architecture and the software stack design. The main contributions of this paper are:

\squishlist
    \item We propose a two-level instruction dispatch module and a multi-core based hardware resources pool to provide isolated and runtime-programmable hardware resources.
    \item We propose a tiling-based instruction frame package design and two-stage static-dynamic compilation to overcome the heavy online reconfiguration overhead.
    \item The experimental results show that our virtualization solution achieves 1.07-1.69x and 1.88-3.12x higher throughput in the private cloud scenario compared with the static single-core design and static multi-core design, respectively.
\squishend


\section{Related work}


\begin{figure*}[!tp]
    \centering
	\includegraphics[width=17.5cm]{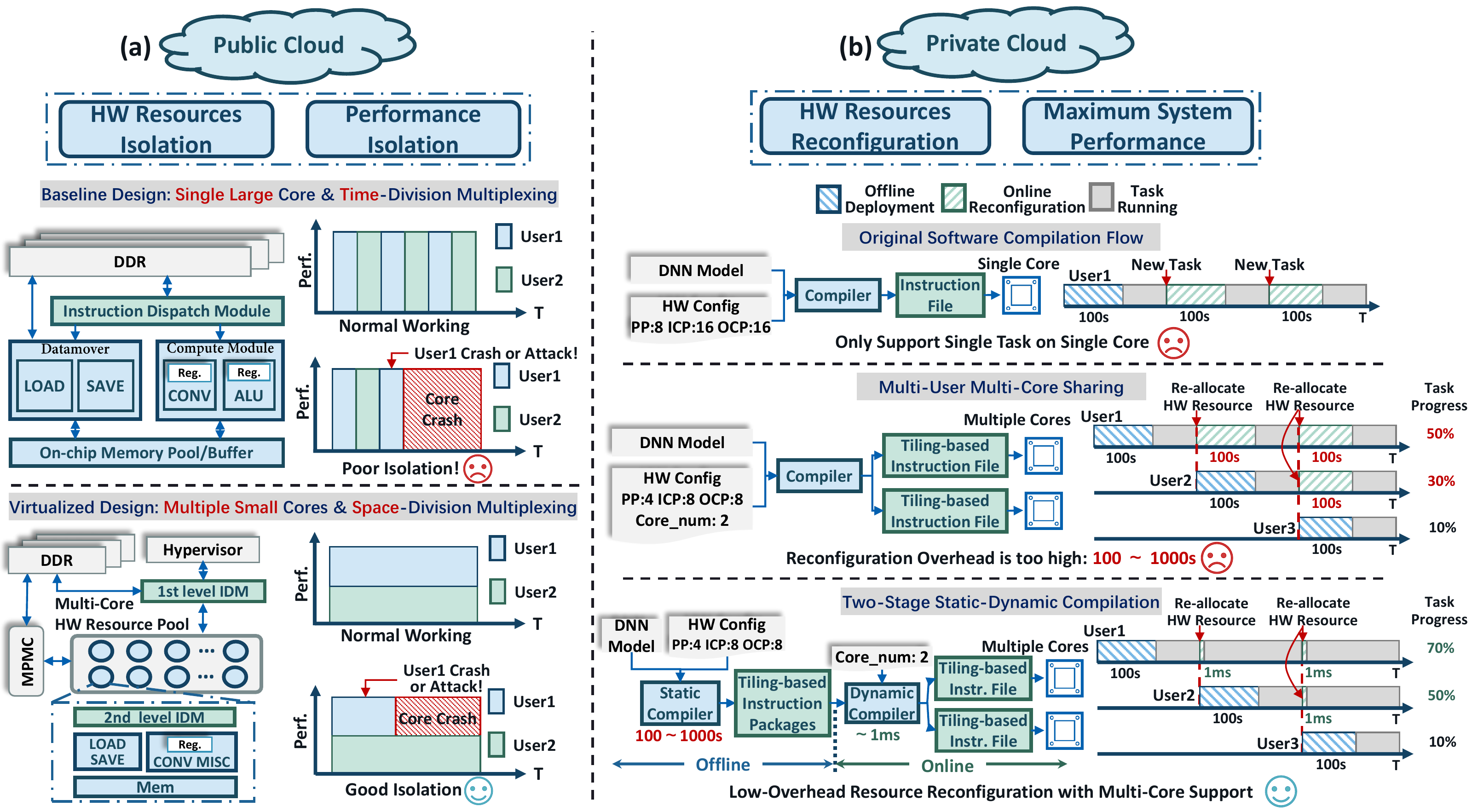}
	\vspace{-10pt}
	\caption{Overall ISA-based virtualization methodology: (a) Virtualized ISA-based hardware architecture design targeted for the public cloud. (b) Low-overhead reconfiguration software compiler design for the private cloud. PP: pixel parallelism. ICP: input channel parallelism. OCP: output channel parallelism.}
	\vspace{-10pt}
	\label{fig:isa-system-method}
\end{figure*}

\subsection{FPGA Virtualization in the cloud}
In order to integrate FPGAs in the cloud, hardware virtualization techniques are required. Vaishnav \emph{et al.}~\citep{vaishnav2018survey} summarized research on FPGA virtualization and classified these work into three categories: resource level, node level, and multi-node level. At the resource level, hardware resources of FPGAs are divided into reconfigurable resources (e.g., logic) and non-reconfigurable resources (e.g., I/Os). For the virtualization of reconfigurable resources on the FPGA chip, a common way is to implement an intermediate overlay architecture~\citep{brant2012zuma, cong2014fully} between high-level software framework and low-level FPGA hardware. While I/O virtualization enables the sharing of hardware resources by different tasks, using the same I/O interface. Node level FPGA virtualization treats an FPGA chip as a computation node, and multiple accelerators on the FPGA chip are used to execute different tasks simultaneously. Chen \emph{et al.} proposed the FPGA virtualization architecture at both resource and node level using partial reconfiguration~\citep{chen2014enabling}. Knodel \emph{et al.} also proposed similar FPGA virtualization architecture~\citep{knodel2017virtualizing}. Dai \emph{et al.} further proposed the scheduling scheme over such partial reconfiguration architecture~\citep{dai2014online}. Multi-node level FPGA virtualization uses multiple FPGAs to provide an acceleration system for tasks. Microsoft proposed Catapult system to accelerate its online Bing search application~\citep{putnam2014reconfigurable}. Baidu also proposed their software-defined FPGA accelerators for deep learning and big data applications in their data center~\citep{ouyang2016sda}. Baidu's XPU~\citep{ouyang2017xpu} is designed for diverse workloads using many tiny ISA-based cores. However, it didn't discuss the details of how to support multiple users and dynamic workload in actual cloud scenarios.

\subsection{DNN Accelerators}
Benefiting from the flexibility, programmability, and energy efficiency of FPGAs, FPGA based accelerators can provide high performance and energy efficiency for deep learning applications~\citep{wang2019deep, guo2019dl}. Recent research on FPGA based DNN accelerators can be divided into two categories: (1) using register-transfer level (RTL) or high-level synthesis (HLS) based templates to map the target DNN model into several computation blocks~\citep{zhang2018dnnbuilder, zhang2018caffeine}. Such design flow requires to regenerate the bitstream file for each input DNN model and reprogram the FPGA. The other approach is (2) introducing a customized ISA~\citep{abdelfattah2018dla, chang2017compiling}. The ISA-based DNN accelerators don't need to reprogram the FPGA, while it can reconfigure its DNN task by reloading the updated instruction files. However, this kind of accelerator requires a carefully optimized compiler~\citep{sharma2016dnnweaver, chen2018tvm} to map DNN models to the underlying hardware architecture efficiently. 

A few prior studies focused on the FPGA-based DNN accelerators in the cloud. Chen \emph{et al.} proposed an automated framework for mapping DNN models to cloud FPGAs~\citep{chen2019cloud}. Han \emph{et al.} proposed a sparse LSTM accelerator to deploy speech recognition applications in the cloud~\citep{han2017ese}. Some technology giants have also proposed DNN accelerator solutions in the cloud. Intel~\citep{aydonat2017opencl} used OpenCL to provide a DNN deployment framework and accelerator in the cloud. Xilinx~\citep{xilinx2018xDNN} designed an ISA-based CNN accelerator targeted at cloud FPGAs. However, previous work mainly focused on optimizing the performance of running the DNN model for the single-task and static-workload scenarios in the cloud-based FPGA, without considering the characteristics of multi-user and dynamic-workload scenarios in the cloud. Microsoft~\citep{fowers2018configurable} proposed a multi-FPGA virtualization system framework to serve the DNN workloads in the cloud. It demonstrated great performance on sharing multiple FPGAs, but did not discuss the case of sharing a single FPGA, which should be also important and require more in-depth research.

From previous work on FPGA virtualization and deep learning accelerators, we can see that few studies focused on both perspectives. Researchers didn't design the DNN accelerators from the perspective of the actual application scenarios in the cloud. Without considering the multi-task and dynamic-workload applications in the cloud, such DNN accelerators are still far away from providing efficient and flexible FPGA virtualization solutions.

\section{ISA-based Virtualization Methodology}\label{sec:isa-based system}
In this section, we will give a brief introduction to our ISA-based virtualization methodology to enable efficient and flexible FPGA virtualization for deep learning inference applications in the cloud. Our virtualization methodology is applicable to any ISA-based DNN accelerator, and we will give a demonstration based on Angel-Eye~\citep{guo2017angel}. As shown in Figure~\ref{fig:isa-system-method}, in order to support both public and private cloud scenarios, both isolation and high performance are required. Thus, we propose virtualized ISA-based hardware architecture design and low-overhead reconfiguration software compiler design, which will be discussed in Section~\ref{sec:Hardware architecture} and~\ref{sec:software compiler}, respectively.

\subsection{Isolation: ISA-based Hardware Architecture with Space-Division Multiplexing}
As shown in the upper part of Figure~\ref{fig:isa-system-method}(a), the current single-core design is an ISA-based CNN accelerator with a TDM method to support multi-user scenario. When a user crashes or performs a malicious attack, the running tasks of other users will be forced to stop due to the crash of the same physical computing core shared by multiple users. This potential security issue is contrary to the requirements for physical resources and performance isolation in the public cloud. In contrast, SDM can satisfy the isolation requirements in the public cloud by letting different users monopolize different physical resources. We proposed a \textbf{multi-core hardware resources pool} technique to support the SDM based multi-user scenario, by dividing a single large core into multiple small cores. Thus, each user can monopolize a given number of small cores in the hardware resources pool (HRP) to obtain a safe and undisturbed operating environment.


One problem brought by SDM is how to achieve similar single-task performance to the single large core by multi-core sharing under the same hardware resources. The performance loss of multi-core sharing mainly comes from the overhead of communication and synchronization. Since a small core can achieve a better utilization rate than a large core, we can minimize the performance loss by designing an appropriate tiling method to minimize the communication and synchronization overhead of multi-core sharing while ensuring a good utilization rate. As shown in Figure~\ref{fig:isa-system-method}(b), we introduce a \textbf{tiling-based instruction package design} to enable multi-core sharing with negligible performance loss in the software level. The basic idea is to tile the output feature map in a workload-balanced manner among all the allocated cores, with an accurate latency simulator for the workload prediction. Besides, we introduce a \textbf{two-level instruction dispatch module} to handle multi-user task scheduling and single-task multi-core sharing synchronization control at the hardware level.

\subsection{High Performance: Low-overhead Reconfiguration Software Compiler Design}

With a pool of virtualized hardware resources, the mapping from the DNN models to the ISA-based multi-core DNN accelerators must be recompiled on every resources re-allocation. If the re-compilation is too slow, it could lead to severe tail latency of serving multi-user DNN requests in the private cloud scenario. To minimize the overhead of online reconfiguration, we propose a \textbf{two-stage static-dynamic compilation}, where the original compilation flow is divided into two parts: static compiler and dynamic compiler, as shown in Figure~\ref{fig:isa-system-method}(b). The former generates the tiling-based instruction frame packages (IFPs) during the offline deployment stage based on the input DNN model and the hardware configuration of the basic shareable unit. The latter generates the tiling-based instruction files for each user by re-allocating the pre-generated tiling-based IFPs to each core based on the re-allocated hardware resources. We demonstrate that such a compilation process can reduce the overhead of online reconfiguration to about 1ms by only recompiling the light-weight runtime information, thereby ensuring that the task progress of each user is hardly affected even in the case of frequent dynamic reconfiguration.

\begin{figure}[!tp]
\centering
\includegraphics[width=0.45\textwidth]{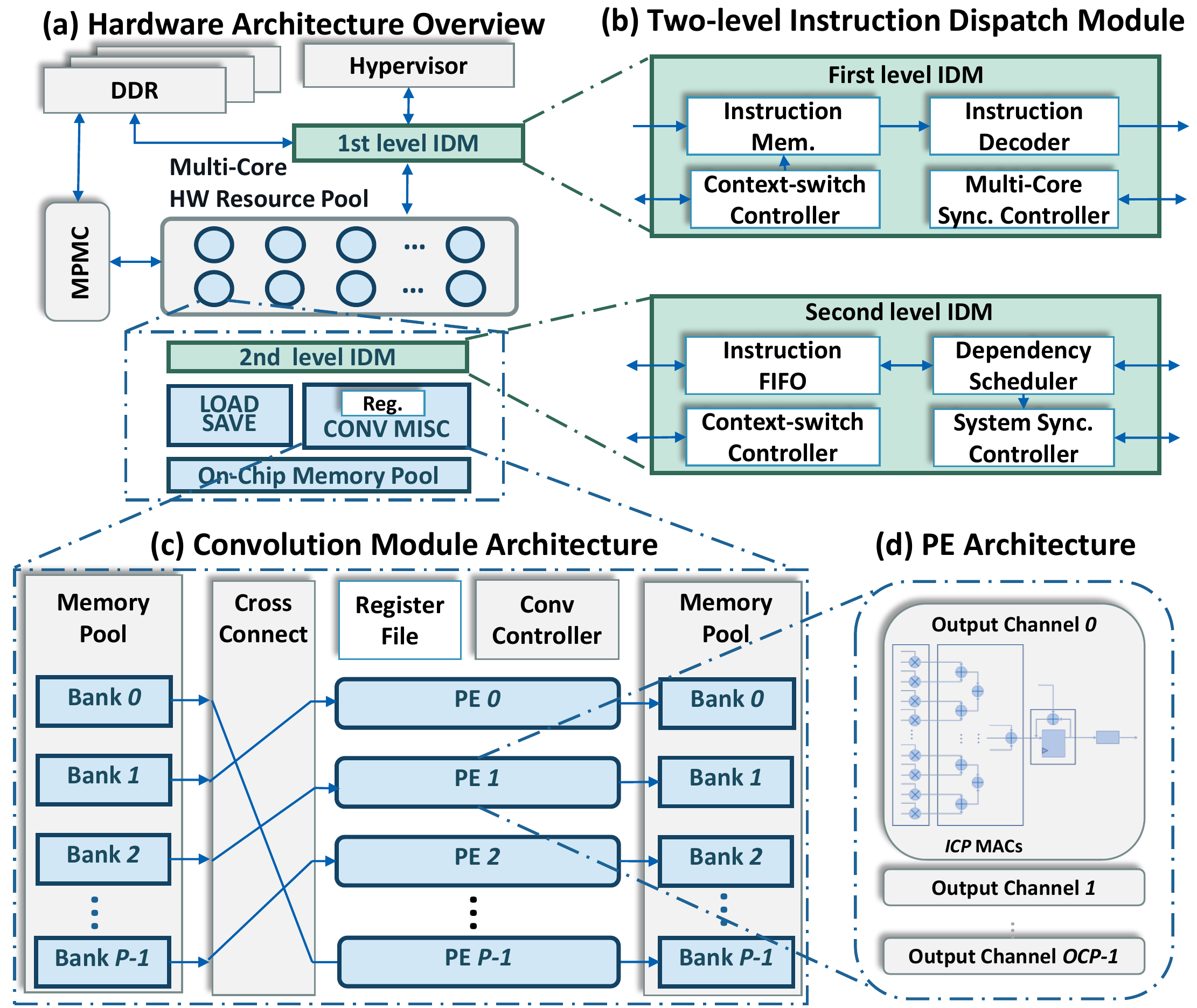}
\vspace{-10pt}
\caption{The hardware design for virtualization.}
\vspace{-10pt}
\label{fig:hardware-design}
\end{figure}

\section{Hardware Architecture}\label{sec:Hardware architecture}


\subsection{Baseline Design}\label{sec:hw-base}
Our ISA-based DNN accelerator baseline design is based on Angel-Eye~\citep{guo2017angel}, as shown in Figure~\ref{fig:isa-system-method}(a). It contains an instruction dispatch module (IDM) and four instruction-related modules, namely the convolution computation module (CONV), the non-convolution computation module (MISC), and the datamover module (LOAD/SAVE). Figure~\ref{sec:Hardware architecture}(c) and (d) show the hardware architecture of the CONV module and the processing element (PE), respectively. Its computation parallelism is shown as below:

\vspace{-10pt}
\begin{equation}
    \label{eq:parallelism}
    Parallelism = 2 \cdot PP \cdot ICP \cdot OCP \quad (OPs/cycle)
\end{equation}
where $PP$, $ICP$, and $OCP$  represent the parallelism along the pixel, the input channel, and the output channel dimensions of the feature map, respectively. Corresponding to the CONV module architecture, $PP$ equals to the number of PE $P$, meaning that each PE completes the computation of one pixel of feature map per cycle. Inside each PE, there are a total of $OCP$ parallel computing channels, each corresponding to the computation of one output channel. For each computation channel, it can handle the multiply-accumulate (MAC) computation between the feature maps and weights of $ICP$ input channels in each cycle, so the computation parallelism in Equation~\ref{eq:parallelism} needs to be multiplied by two. 


\subsection{Hardware Design for Virtualization}\label{sec:hardware-res}

\subsubsection{Two-level Instruction Dispatch Module}

The main function of the original IDM is to implement instruction distribution and dependency management in a single core. When it comes to the hardware design for virtualization, the shareable units in the multi-core HRP also need to be scheduled for the multi-user support. Thus, we need a two-level instruction distribution module to schedule and manage the hardware resources of two different dimensions.

\textbf{First Level Instruction Dispatch Module}. The first level IDM can be regarded as a task-level scheduler. Its basic components are shown in Figure~\ref{fig:hardware-design}(b). The on-chip instruction memory fetches the instructions from DDR and caches them until the next reconfiguration. The instruction decoder sends the instructions to the second level IDM of the corresponding core according to the core index of each instruction. 

The context-switch controller will record the context information after receiving the reconfiguration signal from the hypervisor. Currently, it supports two context-switch modes: task-level and layer-level switching. The first mode only needs to wait for the implementation of the current inference task and then load new instructions into each core. In the second mode, the context information to be recorded is the DNN layer index of each user. Since our baseline design works in a layer-by-layer manner, intermediate data such as feature map will be written back to DDR when a layer calculation is finished, so there is no need to record it as context information. Next, the context-switch controller will load the updated instructions and layer index to all cores of each user, so that the computing cores can continue to calculate from the next layer.

The multi-core synchronization controller is to manage the layer-wise multi-core synchronization, which will be discussed in Section 5.2.2. The hypervisor configures this module to control which cores need to be synchronized. It will read in the $sync\_local$ signals of all cores belonging to each user. Only when all $sync\_local$ signals are valid, will it send a valid $sync\_global$ signal to each core, so that these cores can start the calculation for the next layer.

\textbf{Second Level Instruction Dispatch Module}. The second level IDM can be regarded as a module-level scheduler inside each core. As shown in Figure~\ref{fig:hardware-design}(b), the lower two sub-modules are specifically designed for virtualization. The context-switch module restarts the computation based on the context information recorded by the first level IDM in the online reconfiguration stage. The system synchronization controller generates the $sync\_local$ signal when the $System$ instruction for synchronization of the current layer satisfies the dependency constraints, meaning that the computation of the current layer finishes. Then it suspends the scheduler to stop dispatching instructions until it receives a valid $sync\_global$ signal for the next layer to start.



\subsubsection{Multi-Core Hardware resources Pool}
The basic idea of virtualized HRP is to divide the single large core in the baseline design into multiple small cores to provide isolated and runtime-programmable hardware resources. Here we will discuss the implementation techniques to ensure hardware resources isolation and performance isolation to meet the requirements in the public cloud.

\textbf{Hardware resources Isolation}. The shareable on-chip physical resources include on-chip memory, on-chip bandwidth, and PE arrays consisting of DSPs. The multi-core design isolates these physical resources fundamentally. Each core cannot access the memory or computing resources of others. And each core can only perform basic synchronous interaction with other cores through the first level IDM. Other important physical resources are off-chip DDR memory and bandwidth. As FPGAs in the cloud, such as Xilinx VU9P and U200, usually have up to 4 independent DDR banks, the safest way is that each user can monopolize single or multiple DDR chips. However, it limits the maximum number of users that a single FPGA can serve. Sharing a single DDR by multiple users requires proper isolation measures at the operating system level, which is not discussed in this paper.

\textbf{Performance Isolation}. The performance loss caused by multiple users mainly comes from the competition for the same physical resources. According to the above discussion, competition can only occur when multiple users share a single DDR. Since each DDR has only a single 512-bit data port, if the total demand for the memory bandwidth of multiple users exceeds 512 bits, performance will inevitably be deteriorated due to the competition among multiple data ports. Thus, a basic hardware restriction is that the total bit width of multi-user data ports cannot exceed the data bit width of a single DDR data port. In addition, a well-designed arbiter is needed to ensure that performance crosstalk between multi-user data ports is minimized, which has been well studied and tested in recent years~\citep{axi-inter}. By utilizing the techniques discussed, we can ensure that the FPGA virtualization design based on the multi-core HRP achieves good performance isolation in the multi-user public cloud scenarios.  

\begin{figure*}[tphb]
\centering
\includegraphics[width=17.5cm]{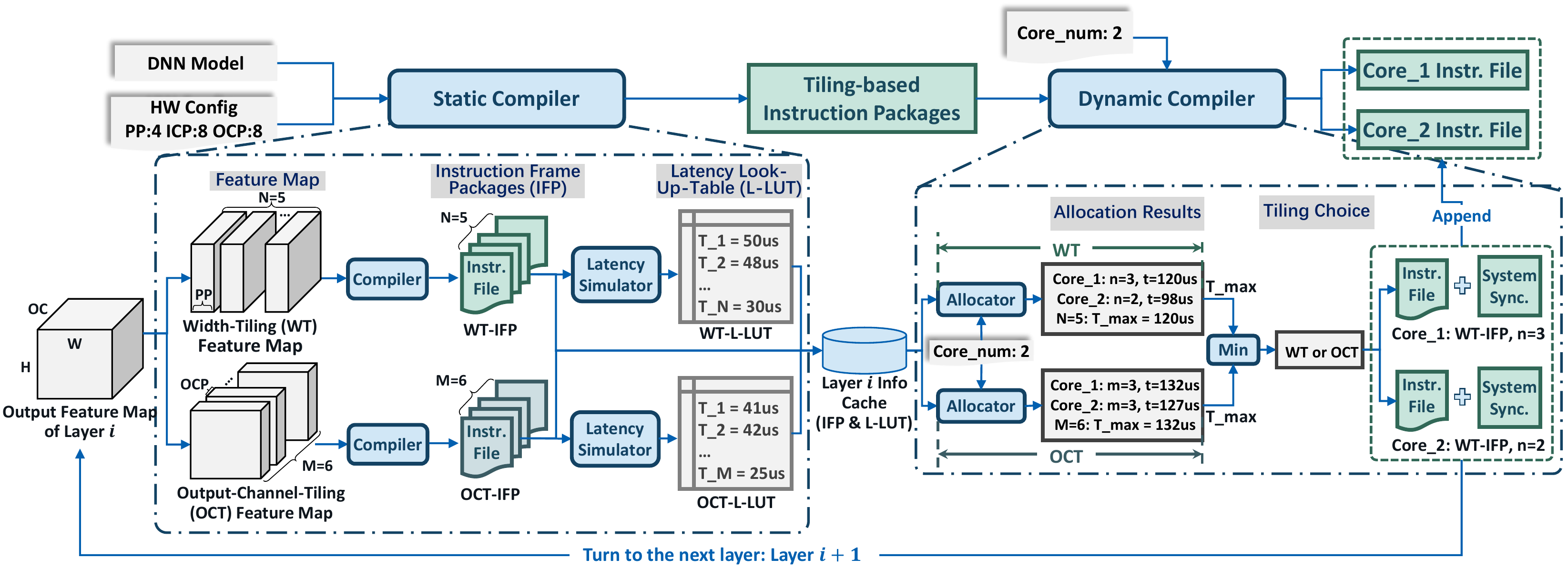}
\vspace{-10pt}
\caption{A complete compilation workflow for virtualization, including the static compiler and the dynamic compiler.}
\vspace{-10pt}
\label{fig:runtime-compiler}
\end{figure*}

\section{Software Compiler}\label{sec:software compiler}

\subsection{Baseline Design}

The core set of ISA is composed of $System$, $Load$, $Save$, $Convinit$, $Conv$, $Poolinit$, and $Pool$. These instructions correspond to the four major functional blocks of the hardware architecture: LOAD, SAVE, CONV, and MISC. The $Load$/$Save$ instructions implement data interaction between DDR and memory pool. The $Convinit$/$Poolinit$ instructions implement layer parameter configuration for the register files of the CONV and MISC modules. Each $Conv$ and $Pool$ instruction correspond to the computation of $PP$ lines and one line of the output feature map with all output channels, respectively. The $System$ instruction is the finish signal of one inference task. In addition, all instructions need to contain dependency information to ensure that hardware and data dependencies are met.



\subsection{Compiler Design for Virtualization}\label{sec:soft-runtime-res}

\subsubsection{Static Compilation}\label{sec:soft-runtime-static}

As shown in Figure~\ref{fig:runtime-compiler}(a), the static compiler generates tiling-based IFPs in a layer-by-layer manner and applies a latency simulator to obtain a latency look-up-table (LUT), which records the latency of each IFP. The tiling-based IFPs and latency LUT are cached for online reconfiguration.


\textbf{Tiling-Based Instruction Frame Packages}. In order to support multi-core sharing of a single task, the output feature map of each layer must be tiled into several independent sub-tiles for parallel computing. Since the compiler generates $Conv$ instructions along the height dimension, tiling along the height dimension will result in a complex dependency relationship between the IFPs. Thus, we choose the output channel and width dimensions as our target tiling dimensions to generate independent IFPs.

The additional overhead introduced by multi-core sharing is that each core may need to load the same data from DDR to the on-chip memory, thus reducing the overall data reuse efficiency. For tiling along the output channel, it can be regarded as weight parallelization, meaning that each core will load a different part of weights but the same input feature map. In contrast, tiling along the width will load a different part of input feature maps but the same weights. Since the size of the input feature map and weights of each layer are quite different, different layers will have different preferences for the two tiling methods to obtain a better latency. Thus, we can minimize the performance loss of multi-core sharing by choosing the proper tiling method for different layers.

\textbf{Latency Simulator}. We design a fast latency simulator to obtain a cycle-accurate latency evaluation of each tiling-based IFP. For the $Conv$ instruction, we estimate its latency based on the computation amount and the computation parallelism:

\vspace{-10pt}
\begin{equation}
    \label{eq:latency-conv}
    t = \frac{Channel_{in}Channel_{out}}{ICP*OCP}*Width_{out}*Kernel_{w}*Kernel_{h}*T
\end{equation}
where $T$ denotes the clock cycle. For the instructions for data movement including $Load$ and $Save$, their latency can be estimated using the total length of data and bandwidth:

\vspace{-10pt}
\begin{equation}
    \label{eq:latency-load}
    t = \frac{Length_{data}}{BandWidth*eff}
\end{equation}
Where $eff$ is a parameter of bandwidth efficiency. Then, a directed acyclic graph $ G (V, E) $ according to the data and hardware dependencies of the instructions are set up, where $V$ represents each of the instructions in the IFP, and $E$ stores dependencies between instructions. By traversing the entire graph $G$, we can get the latency estimation of the IFP and store it into a latency LUT.

\subsubsection{Dynamic Compilation}\label{sec:soft-runtime-dynamic-compiler}
Figure~\ref{fig:runtime-compiler}(b) shows the workflow of the dynamic compiler. During each online reconfiguration, the dynamic compiler generates the optimized instruction files in a layer-by-layer manner. Firstly, the dynamic compiler fetches the latency LUTs of the two tiling methods from the cache. Secondly, the allocator will find the optimal allocation scheme for multi-core sharing to minimize the latency of the current layer, according to the number of re-allocated cores. Then, the dynamic compiler chooses the tiling method with minimal latency as the target strategy for the current layer. The dynamic compiler will take the corresponding IFPs from the cache, combine it into multiple instruction sequences according to the optimal allocation scheme, and add a synchronization $System$ instruction at the end of each instruction sequence. The dynamic compiler repeats this process until the instructions of all layers are generated.


\begin{figure*}[htbp]
\centering
\includegraphics[width=0.95\textwidth]{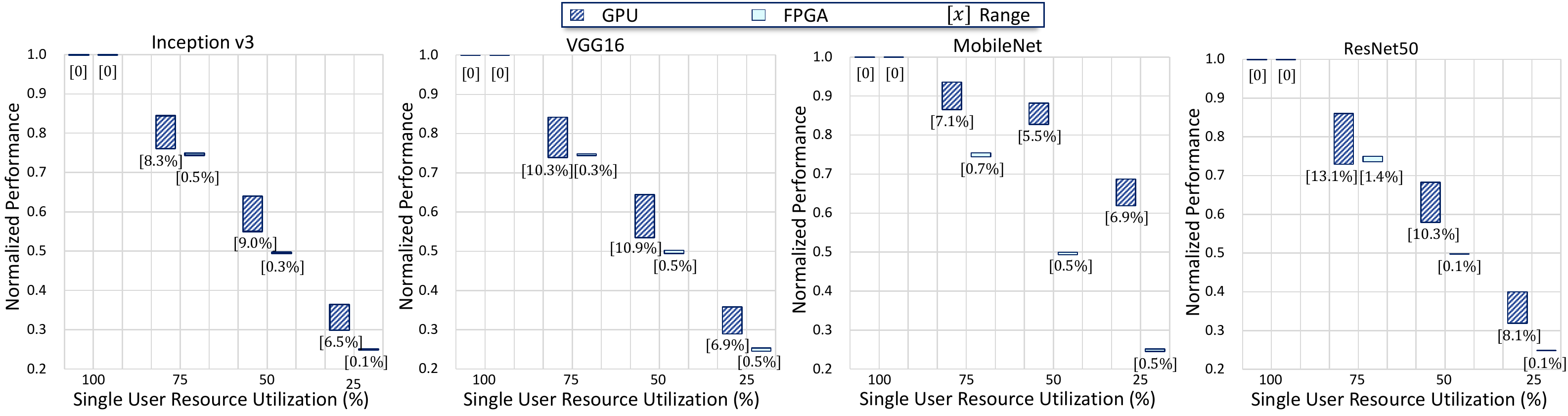}
\vspace{-10pt}
\caption{Performance isolation evaluation of our FPGA virtualization design and GPU virtualization design. The performance deviation of a single user with different hardware resources under multi-user scenario is evaluated.}
\vspace{-10pt}
\label{fig:exp-isolation}
\end{figure*}

\textbf{Workload-Balanced Instruction Allocator}. For each layer, given the number of tiling-based IFPs $N$ and the number of allocated cores $Num_{kernel}$, we need a workload-balanced IFP allocation method $Alloc$. This problem can be modeled as an optimization problem as below:

\vspace{-15pt}
\begin{gather}
    \label{eq:optimizer}
    \mathop{\arg\min}_{Alloc}{\max_{k=1}^{M}{\sum_{i=1}^{N}{Alloc(i,k)T(i)}}} \\
    \sum_{k=1}^{Num_{kernel}}{Alloc(i,k)} = 1, \forall i \in \{1,\dots,N\} \\
    Alloc(i,k)\in \{0,1\}, \forall i \in \{1,\dots,N\}, \forall k \in \{1,\dots,M\}
\end{gather}
where $Alloc(i,k) = 1$ denotes the $i$-th IFP is allocated to the $k$-th core. $T(i)$ denotes the latency of the $i$-th IFP. This optimization problem can be solved quickly using classic dynamic programming methods. Thus, we can use this allocator to quickly get the workload-balanced allocation method of the two tiling methods in each layer.

\textbf{Layer-Wise Multi-Core Synchronization}.
Since the computing workload of each core in the same layer is not exactly the same, a synchronization mechanism must be introduced to ensure correct data dependencies. We add a synchronization bit to the function field of the $System$ instruction. When each core runs to the synchronization $System$ instruction, it stops running and enters the synchronization waiting state. Only when all the cores of a user have run to the $System$ instruction of the current layer, can the calculation of the next layer be started.


\subsubsection{Context Switching Analysis} 
The task switching of the ISA-based DNN accelerator is realized through the regeneration and reloading of new instruction files. Thus, its context switching cost is mainly composed of two parts: (1) $T_{recompile}$, i.e., the compiling time to regenerate the instruction files; (2) $T_{transfer}$, i.e., the time of sending the new instruction files from the hypervisor to the DNN accelerator. The overall context switching cost $T_{context}$ can be estimated as followed:


\vspace{-10pt}
\begin{equation}
    \label{eq:switch-cost}
    T_{context} = T_{recompile} + T_{transfer}
\end{equation}

\section{Experiments}\label{sec:exp}

\subsection{Experiment Setup}
\textbf{Hardware Platform}. We evaluate our virtualized design on two different hardware platforms. For the private cloud scenario, we set up a local machine with Xilinx Alveo U200 FPGA and an Intel Xeon 4210 CPU running at 2.2GHz. For the public cloud scenario, we choose the f3 instance in Aliyun, which contains a Xilinx VU9P FPGA and an Intel Xeon Platinum 8163 CPU running at 2.5GHz. The GPU platform for isolation evaluation consists of an Nvidia Tesla V100 GPU and an Intel Xeon Gold 6132 CPU running at 2.60GHz.

\textbf{Software Environment}. We use Xilinx SDAccel 2018.3 for hardware synthesis and software deployment in both the local machine and the Aliyun cloud environment. The software compiler design for virtualization is implemented in Python, and the host application is developed using C++ with OpenCL APIs provided by SDAccel. To enable performance isolation on GPU, we use Nvidia CUDA MPS (Multi-Process Service)~\citep{nvidia2019mps} by setting the environment variable \texttt{CUDA\_MPS\_ACTIVE\_THREAD\_PERCENTAGE}.  

\textbf{CNN Models}. We evaluate four CNN models with input image size of $224\times224$: VGG16~\citep{simonyan2014very}, ResNet50~\citep{resnet}, Inception v3~\citep{szegedy2016rethinking}, and MobileNet~\citep{howard2017mobilenets}. We use caffe~\citep{jia2014caffe} pre-trained models for FPGA and tensorflow~\citep{abadi2016tensorflow} 1.12 official benchmarks for GPU.

\textbf{CNN Accelerator Configuration}. For the baseline design, we set up two different configurations: a static single large core design with the parallelism of 8192 and a static multi-core design with 16 small cores, each of which has a parallelism of 512. The virtualized design has the same configuration with the static multi-core design of $16 \times 512$ parallelism. All the accelerators run at 300MHz. The memory bandwidth of each small core is 128 bits, and every four small cores share a single DDR. The static single large core has access to four DDR banks for a fair comparison.

\begin{table}[t]
  \centering
  \scriptsize
  \caption{Hardware resources utilization of the static baseline design and virtualized design on Xilinx U200 and VU9P FPGA using SDAccel 2018.3.}
  \vspace{-10pt}
    \begin{tabular}{c|c|ccccc}
    \hline
    FPGA  & Implementation & LUT   & FF    & BRAM  & URAM  & DSP \bigstrut\\
    \hline
    \multirow{4}[2]{*}{U200} & User Budget & 891205 & 1960162 & 1153  & 960   & 6824 \bigstrut[t]\\
          & Static Single-Core (8192) & 242135 & 232588 & 235   & 168   & 2048 \\
          & Static Multi-Core (16$\times$512) & 418282 & 389777 & 395   & 307   & 2048 \\
          & Virtualized Design (16$\times$512) & 435710 & 401832 & 416   & 320   & 2048 \bigstrut[b]\\
    \hline
    \multirow{4}[2]{*}{VU9P} & User Budget & 891201 & 1960522 & 1153  & 960   & 6824 \bigstrut[t]\\
          & Static Single-Core (8192) & 241580 & 231799 & 235   & 168   & 2048 \\
          & Static Multi-Core (16$\times$512) & 418901 & 390125 & 395   & 307   & 2048 \\
          & Virtualized Design (16$\times$512) & 435722 & 401277 & 416   & 320   & 2048 \bigstrut[b]\\
    \hline
    \end{tabular}%
  \label{tab:resource}%
\end{table}%

\begin{table}[t]
  \centering
  \scriptsize
  \vspace{-10pt}
  \caption{Compilation and context switching cost with the number of re-allocated cores as 1, 2, 4, 8, and 16.}
  \vspace{-10pt}
    \begin{tabular}{c|c|c|c|c}
    \hline
    \multirow{2}[2]{*}{CNN Model} & Static & Dynamic & \multirow{2}[2]{*}{T\_{transfer} (ms)} & Context Switch \bigstrut[t]\\
          & Compilation (ms) & Compilation (ms) &       & Cost (ms) \bigstrut[b]\\
    \hline
    VGG16 & 44795.6 & 0.4-0.65 & 0.05-0.18 & 0.45-0.83 \bigstrut\\
    ResNet50 & 46807.9 & 0.86-1.06 & 0.03-0.15 & 0.89-1.21 \bigstrut\\
    Inception v3 & 34882.9 & 1.06-1.5 & 0.06-0.20 & 1.12-1.70 \bigstrut\\
    MobileNet & 14657.6 & 0.53-0.67 & 0.03-0.15 & 0.56-0.82 \bigstrut\\
    \hline
    \end{tabular}%
    \vspace{-10pt}
  \label{tab:switch-cost}%
\end{table}%

\subsection{Resources Utilization and Context Switch}

\subsubsection{Hardware resources Utilization}
As shown in Table~\ref{tab:resource}, the static multi-core design utilizes nearly twice more logic and memory resources than the static single-core design with the same DSP resources. This is because there are multiple copies of each module in the multi-core design, while some modules, such as datamover and cross-connection module, can be reused in the single-core design. As for our virtualized multi-core design, it introduces about 1\% logic and memory resources overhead compared to the static multi-core design, which mainly comes from the two-level IDM design. The main resources utilization of the current FPGA virtualization design is logical resources, which occupy nearly 50\% of LUTs and FFs. This means that we can reduce resources utilization by optimizing the logic and data paths of the single-core, thereby further scaling the virtualization design up to a higher degree of parallelism, which will be implemented in our future work.

\begin{figure*}[htbp]
\centering
\vspace{-40pt}
\includegraphics[width=0.95\textwidth]{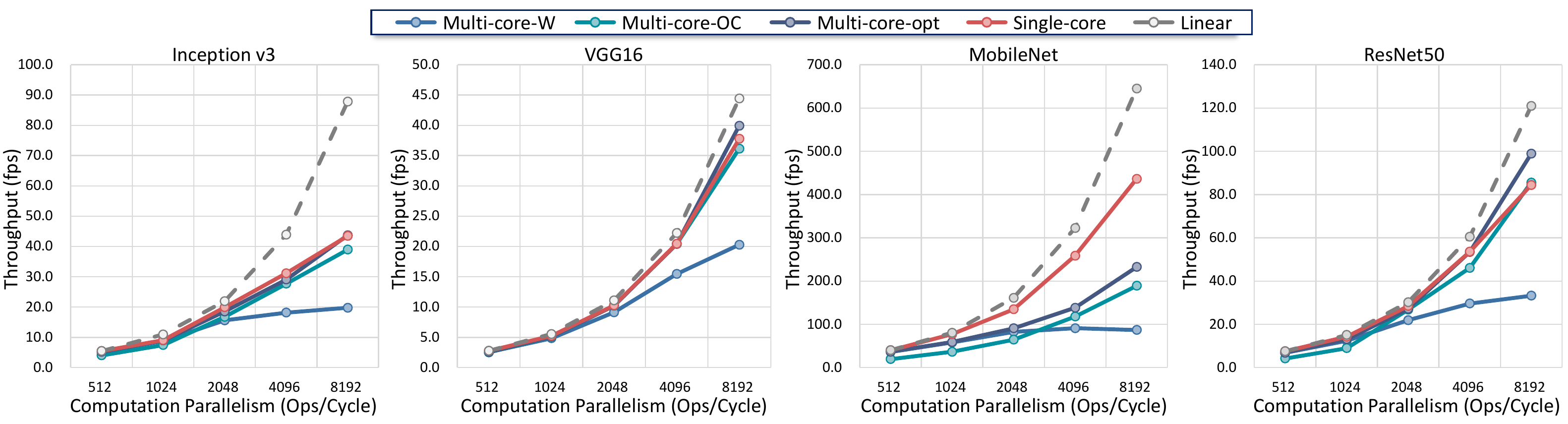}
\vspace{-10pt}
\caption{The single-task throughput of the virtualized multi-core designs with three different tiling strategies (W: width-only, OC: output-channel-only, opt: optimized) and the single-core baseline design under different computation parallelism.}
\vspace{-10pt}
\label{fig:exp-single-task}
\end{figure*}

\begin{figure*}[htbp]
\centering
\includegraphics[width=0.95\textwidth]{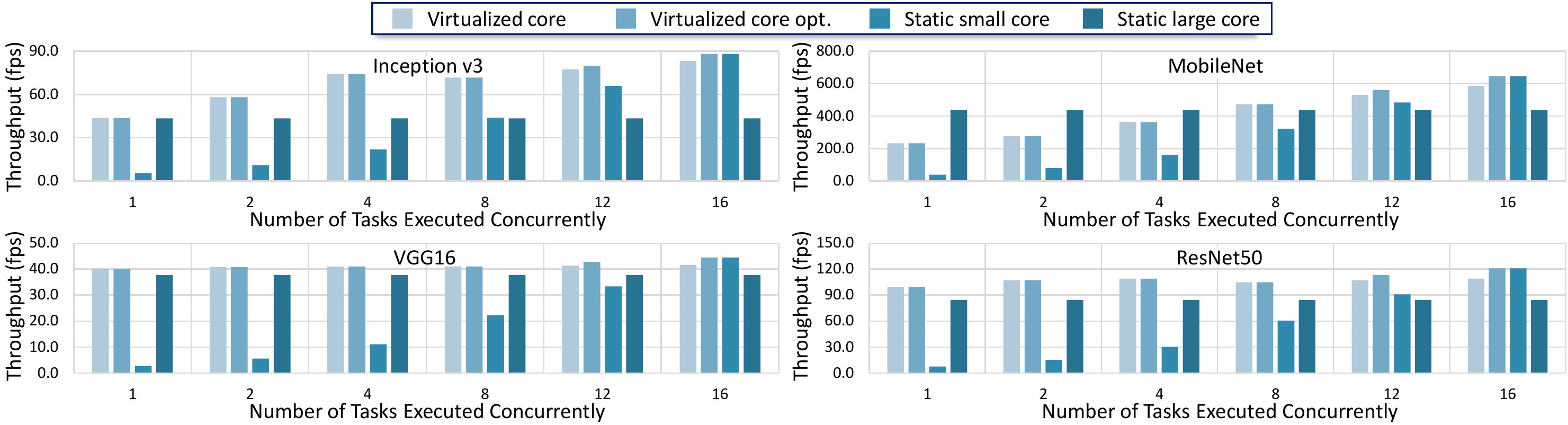}
\vspace{-10pt}
\caption{The multi-task throughput of the virtualized multi-core designs, the static multi-core design, and the single-core baseline design under different workload. (opt: optimization for the case where each core runs a single task.)}
\vspace{-10pt}
\label{fig:exp-multi-task}
\end{figure*}

\subsubsection{Context Switching Cost Analysis}
Table~\ref{tab:switch-cost} shows the compilation and context switching cost of four CNN models with the number of re-allocated cores as 1, 2, 4, 8, and 16. It takes the static compiler 14.7-46.8s to generate the tiling-based IFP during the offline deployment, while the dynamic compilation cost is only 0.4-1.5ms during online reconfiguration. Considering the overhead of transferring the instruction files to the CNN accelerator, the online reconfiguration overhead is limited to 0.45-1.70ms. Since our software compiler design is implemented in Python, it is possible to further reduce the dynamic compilation cost by implementing it using C++ and multi-thread optimization methods.  

\subsection{Performance}
\subsubsection{Evaluation on Isolation}
Our FPGA virtualization design enables the hardware resources isolation on FPGA, which contributes to the performance isolation of different users in the public cloud. We compare our FPGA virtualization design on VU9P FPGA with GPU in this section. In the evaluation, we set the maximum number of users to 4 and give one user fixed resources $x$ (100\%, 75\%, 50\%, and 25\%), and then adjust the remaining users to occupy $1-x$ resources in different proportions. We can get the maximum and minimum performance of a user with different fixed resources $x$.

As shown in Figure~\ref{fig:exp-isolation}, when a user monopolizes all resources, there is no performance deviation. When the resources occupied by a single user are 75\%, 50\%, and 25\%, GPU virtualization solution has performance deviations of 7.1-13.1\%, 5.5-10.9\%, and 6.5-8.1\%, while our FPGA virtualization design limits the performance deviation within 1\%. This shows that our FPGA virtualization solution can achieve better isolation than GPU, and it can well meet the requirements for isolation in the public cloud. A single user running MobileNet on GPU will get a non-linear relationship between performance and resources due to the fact that a small DNN model such as MobileNet cannot fully utilize the computing resources of Tesla V100 GPU.

\subsubsection{Evaluation on Single-Task Throughput}
As shown in Figure~\ref{fig:exp-single-task} and Table~\ref{tab:perf-single-task-resnet50}, the virtualized multi-core design with width-only tiling method has better throughput when there are fewer cores, while the output-channel-only tiling method is better with more cores. Both of these two tiling methods result in a large performance loss of 19.95\% and 30.26\% on average when compared to the single-core baseline design on ResNet50. By considering the impact of both tiling methods in our two-stage static-dynamic compilation process, we can obtain optimized multi-core instructions with only 1.12\% performance loss on average. We can get similar results on inception v3 and VGG16, the optimized multi-core performance loss of which are 0.95\% and 3.93\% on average, respectively.

\begin{table}[t]
  \centering
  \scriptsize
  \caption{Single-task performance on ResNet50. (W: width-only, OC:output-channel-only, opt: optimized)}
    \begin{tabular}{c|ccccc|c}
    \hline
    \multicolumn{1}{c}{} & \multicolumn{5}{c|}{Throughput (fps)} & Average \bigstrut\\
\cline{1-6}    Parallelism & 512   & 2$\times$512  & 4$\times$512  & 8$\times$512  & 16$\times$512  & Loss \bigstrut\\
    \hline
    multi-core-W & 6.8   & 12.4  & 21.9  & 29.6  & 33.3  & 30.26\% \bigstrut\\
    multi-core-OC & 4.2   & 9.0   & 26.8  & 46.1  & 85.5  & 19.95\% \bigstrut\\
    multi-core-opt & 6.8   & 13.1  & 27.2  & 53.5  & 98.9  & 1.12\% \bigstrut\\
    \hline
    single-core & 7.6   & 14.3  & 28.5  & 53.6  & 84.4  & 0 \bigstrut\\
    \hline
    linear & 7.6   & 15.1  & 30.2  & 60.5  & 120.9  & \textbackslash{} \bigstrut\\
    \hline
    \end{tabular}%
    \vspace{-20pt}
  \label{tab:perf-single-task-resnet50}%
\end{table}%

It should be noted that the average performance loss of the optimized multi-core design on MobileNet is up to 31.64\% as shown in Figure~\ref{fig:exp-single-task}. The reason behind this is that MobileNet is a typical small CNN model, and the ratio of its parameter amount to the computation amount is significantly larger than the other three CNN models, so its demand for memory bandwidth is much greater. It means that the performance of the single small core with a parallelism of 512 and the memory bandwidth of 128 bits is severely limited by the memory bandwidth, making the overall performance of virtualized multi-core design deteriorated heavily. We verify the correctness of our hypothesis through simulation experiments by doubling the memory bandwidth of multi-core and single-core designs. The simulation results show that the average performance loss of the optimized multi-core design on MobileNet is reduced to 5.33\%, which further demonstrates that FPGA-based CNN accelerators are memory bandwidth limited.

\subsubsection{Evaluation on Multi-Task Throughput}
We evaluate the throughput of the virtualized multi-core design, static multi-core design, and static single-core design in the multi-task private cloud scenario by changing the number of concurrent tasks. As shown in Figure~\ref{fig:exp-multi-task}, the left half of each CNN model is a low-workload situation, where static multi-core design cannot fully utilize FPGA resources, resulting in poor performance. The right half is a high-workload situation, where a static single-core design has poor performance due to non-linearity, as shown in Figure~\ref{fig:exp-single-task}. While our virtualized multi-core design can achieve optimal performance in any situation. 

For the case of 16 concurrent tasks, each small core will run a task independently. Since our tiling-based two-stage static-dynamic compilation introduces additional overhead in this case, as shown in Table~\ref{tab:perf-single-task-resnet50}, the throughput of virtualized multi-core design is lower than that of the static multi-core design. To solve this problem, we can use the original compiler to generate the single-core instruction files during the offline deployment, while the dynamic compiler only generates tiling-based instructions for the tasks with the number of allocated cores more than one. Through this optimization, we can ensure the optimal performance in the high-workload situation. The experimental results show that our virtualized multi-core design can achieve 1.07-1.69x and 1.88-3.12x throughput improvement over the static single-core design and static multi-core design, respectively.




\section{Conclusions}
In this paper, we propose an efficient and flexible FPGA virtualization framework for deep learning tasks in the cloud. The virtualized multi-core design enables that hardware resources can be dynamically reconfigured for multi-task and dynamic-workload applications, with 0.95-3.93\% performance degradation for the single-task applications. The proposed two-stage static-dynamic compilation and tiling-based instruction frame package techniques make it possible for fast dynamic reconfiguration in the software level, with the context switching cost limited to $\sim$1 ms (0.45-1.70ms). According to the experimental results\footnote{We provide several available demonstrations for our FPGA virtualization solution on Aliyun f3: https://github.com/annoysss123/FPGA-Virt-Exp-on-Aliyun-f3}, the proposed virtualization method achieves 1.07-1.69x and 1.88-3.12x higher throughput over the static single-core and static multi-core baseline design, respectively.

%
\bibliographystyle{ACM-Reference-Format}
\bibliography{ref}

%
\appendix

\end{document}